\documentclass[12pt]{article}
\usepackage{graphicx}
\usepackage[utf8]{inputenc}

\title{ Functional integration over the factor-space $Diff^{1}_{+}(S^{1})/SL(2,\textbf{R}) $}

\vspace{10 cm}

\author{ Vladimir V. Belokurov$^{1,2}$ and Evgeniy T. Shavgulidze$^{1}$    \\
\\    {\small \em 1. Lomonosov Moscow State University,  Moscow,  Russia }
\\    {\small \em 2. Institute for Nuclear
Research of the Russian Academy of Sciences,  Moscow,  Russia }
\\ {\small  vvbelokurov@yandex.ru ; shavgulidze@bk.ru}}

\date{ \ \ \  }
\begin{document}
\maketitle

\begin{abstract}
An explicit form of the functional measure on
 the factor space $Diff^{1}_{+}(S^{1})/SL(2,\textbf{R})$ is obtained that makes Schwarzian functional integrals calculus simpler and more transparent.

\end{abstract}

\section{Introduction}
\label{sec:intr}

In recent years,
it became clear \cite{(SY)} - \cite{(KitSuh)} that a quantum mechanical model of Majorana fermions with a random interaction (Sachdev-Ye-Kitaev model), the holographic
description of the Jackiw-Teitelboim dilaton gravity, and some other models lead to the same effective  theory with the Schwarzian action
\begin{equation}
   \label{Act1}
  A_{Sch}=-\frac{1}{\sigma^{2}}\int \limits _{S^{1}}\,\left[ \mathcal{S}ch\, \{h,\,t\}+2\pi^{2}\left(h'(t)\right)^{2}\right]dt\,,
\end{equation}
where $h(t)$ is an orientation preserving $(h'(t)>0)$ diffeomorphism of the unit circle $S^{1} $ $
(h\in Diff^{1}_{+}(S^{1}))\,.$ And
$$
\mathcal{S}ch \,\{h,\,t\}=
\left(\frac{h''(t)}{h'(t)}\right)'
-\frac{1}{2}\left(\frac{h''(t)}{h'(t)}\right)^2
$$
is the Schwarzian derivative.

The functional integrals in these theories are integrals over the group  $ Diff^{1}_{+}(S^{1})$ with the measure
\begin{equation}
   \label{Measure}
   \tilde{\mu}_{\sigma}(dh)=\exp\left\{- A_{Sch} \right\}  dh=
\exp\left\{\frac{2\pi^{2}}{\sigma^{2}}\int \limits _{S^{1}}\,\left(h'(t)\right)^{2}\,dt  \right\}\, \mu_{\sigma}(dh) \,,
\end{equation}
where
\begin{equation}
   \label{MeasureS}
   \mu_{\sigma}(dh)=\exp\left\{\frac{1}{\sigma^{2}}\int \limits _{S^{1}}\, \mathcal{S}ch\, \{h,\,t\}\,dt  \right\}  dh
\end{equation}
is a quasi-invariant measure on $ Diff^{1}_{+}(S^{1})$ \cite{(Shavgulidze1978)}, \cite{(Shavgulidze2000)}.

However, the direct functional integration over the measure (\ref{Measure}) may lead to senseless infinite results. The reason is  that the
measure (\ref{Measure}) is invariant under the left action of $SL(2,\textbf{R})$ group on the space $Diff^{1}_{+}(S^{1})\,:$
\begin{equation}
   \label{h}
h=\varphi \circ f\ \ \ (\,h(t)=\varphi(f(t))\,)\,;\ \ \ \ \ \varphi\in SL(2,\textbf{R});\ \ \ h,\,f \in Diff^{1}_{+}(S^{1})\,.
\end{equation}
To prove the invariance
\begin{equation}
   \label{invariance}
   \mathcal{S}ch\, \{h,\,t\}+2\pi^{2}\left(h'(t)\right)^{2}=\mathcal{S}ch\, \{f,\,t\}+2\pi^{2}\left(f'(t)\right)^{2}\,,
\end{equation}
 consider the following realization of the group $SL(2,\textbf{R})$
\begin{equation}
   \label{phi}
  \varphi (f)=\frac{1}{i2\pi}\,\log\,\frac{e^{i2\pi f}+z}{\bar{z}e^{i2\pi f}+1}\ ,
\end{equation}
and use the known property of Schwarzian derivative
\begin{equation}
   \label{property}
   \mathcal{S}ch\, \{\varphi\circ f,\,t\}=\mathcal{S}ch\, \{\varphi,\,f(t)\}\,\left(f'(t)\right)^{2}+\mathcal{S}ch\, \{f,\,t\}\,.
\end{equation}

To obtain finite results for functional integrals,  one should factor out the infinite input of the noncompact group $SL(2,\textbf{R})$, that is,
one should integrate over the quotient space $Diff^{1}_{+}(S^{1})/SL(2,\textbf{R})\,. $

To this end, in \cite{(BShExact)} - \cite{(BShCalculus)}, we proposed first to evaluate regularized functional integrals over the group  $Diff_{+}^{1}(S^{1})\,,$  and then to normalize them to the corresponding integrals over the group $SL(2, \textbf{R})\,.$ In this way, we
evaluated functional integrals for partition function and correlation functions in Schwarzian theories.

In particular, in \cite{(BShCalculus)}, we explicitly evaluated functional integrals  assigning the SYK two-point and four-point correlation functions.
Since in this case, the Markov property is not valid, and all points of the circle $S^{1}$ give a nonzero input into the integrals, neither time-ordered nor out-of-time-ordered four-point correlation function is represented in the form of a product of two two-point correlation functions.

It should be noted that
the results for the SYK correlation functions defined as functional integrals over $Diff^{1}_{+}(S^{1})/SL(2,\textbf{R}) $ were obtained for the first time in \cite{(BShCalculus)}.

In the papers \cite{(BAK1)}, \cite{(MTV)},  as well as in  more recent ones studying correlation function in two-dimensional gravity \cite{(KitSuh2)}, \cite{(Yang)},
 \cite{(IPVW)},  the problem of functional integration was reduced to that in the 1D Liouville model or that in the theory of a non-relativistic particle on the hyperbolic upper-half plane placed in a constant magnetic field.
Thereby, the authors of these papers evaluated correlation functions given as the functional integrals over $Diff^{1}_{+}(\textbf{R})/P \,.$\footnote{
 $SL(2, \textbf{R})$ is not a subgroup of the group  $Diff^{1}_{+}(\textbf{R})$ (see, e.g., \cite{(Lang)}).
But the group  $ P $ consisting of transformations $f\rightarrow af+b$ $( P\subset SL(2, \textbf{R}) ) $ is a non-compact
subgroup of  $Diff^{1}_{+}(\textbf{R})\,.$}
It is quite natural that functional integrations over the different integration spaces ($ Diff^{1}_{+}(S^{1})$ and $Diff^{1}_{+}(\textbf{R})$) corresponding to different quantum theories give different results for the same integrand.

There are also some other schemes of dealing with Schwarzian functional integrals
(see, e.g., \cite{(SW)}, \cite{(BShUnusual)}, \cite{(BShPolar)}).
However in this paper, we develop  an approach to a direct functional integration and
find an explicit form of the functional measure on
 the quotient space $Diff^{1}_{+}(S^{1})/SL(2,\textbf{R})\,.$

In section \ref{sec:renorm}, we review the approach to functional integration developed in \cite{(BShExact)} - \cite{(BShCalculus)}.
We demonstrate how using
quasi-invariance of the measure (\ref{MeasureS}) under the group $ Diff^{3}_{+}$, one can evaluate regularized functional integrals over the group $Diff^{1}_{+}(S^{1})\,, $
and then normalize them to the corresponding regularized integrals over the group $SL(2,\textbf{R})\,. $

In section \ref{sec:factor}, we propose another approach to functional integration over the factor-space.
We factorize the measure on the group of diffeomorphisms   $Diff^{1}_{+}(S^{1}) $
and explicitly construct the measure on the factor-space
 $Diff^{1}_{+}(S^{1})/SL(2,\textbf{R}) \,.$

\section{Normalization of functional integrals over $Diff^{1}_{+}(S^{1}) $}
\label{sec:renorm}

First we recall the scheme of a normalization of functional integrals of the form
\begin{equation}
   \label{FI3}
J=\frac{1}{\sqrt{2\pi}\sigma}\int\limits_{Diff^{1}_{+} (S^{1}) }\,\Psi(h)\,
 \tilde{ \mu}_{\sigma} (dh)\,.
\end{equation}

We regularize (\ref{FI3}) as $(\alpha<\pi)$
\begin{equation}
   \label{FI3A}
J^{\alpha}=\frac{1}{\sqrt{2\pi}\sigma}\int\limits_{Diff^{1}_{+} (S^{1}) }\,\Psi(h)\,
 \exp\left\{\frac{2\alpha^{2}}{\sigma^{2}}\int \limits _{S^{1}}\,\left(h' (t)\right)^{2}dt  \right\}\, \mu_{\sigma} (dh)\,,
\end{equation}
and evaluate the regularized integral using the quasi-invariance of the measure (see below).

Generally, the functional integrals (\ref{FI3A})  converge for $ 0< \alpha < \pi \,, $ and diverge for $ \alpha = \pi \,. $

To obtain finite results for functional integration,
we normalize (\ref{FI3A}) to the corresponding integrals over the group $SL(2, \textbf{R})\,.$
Thus, we define functional integrals over the quotient space $Diff^{1}_{+}(S^{1})/SL(2,\textbf{R}) $ as
the normalized functional integrals
\begin{equation}
   \label{JR}
J^{N}=\lim \limits_{\alpha\rightarrow\pi - 0}\  \frac{J^{\alpha}}{ \int \limits _{SL(2, R)}\,\Psi\left(\varphi_{z}\right)\,
 \exp\left\{\frac{2\alpha^{2}}{\sigma^{2}}\int \limits _{S^{1}}\left(\varphi'_{z} (t)\right)^{2}dt  \right\}\,d\nu_{H}}\,,
\end{equation}
where $\varphi_{z}\in SL(2, \textbf{R})$ and $\ d\nu_{H}\ $ is the invariant Haar measure on $SL(2, \textbf{R})\,.$

For $SL(2, \textbf{R})$-invariant integrands $ \Psi(h)$ (it is the case of SYK partition function \cite{(BShExact)} and SYK correlators \cite{(BShCalculus)}),  the denominator is proportional to
the regularized volume of the group $SL(2,\textbf{R})$  explicitly evaluated in \cite{(BShCorrel)}:
$$
V^{\alpha}_{SL(2, R)}=\int \limits _{SL(2,\,R)}\exp \left\{-\frac{2\left[ \pi^{2}-\alpha^{2}\right]}{\sigma^{2}}\,\int \limits _{0}^{1}(\varphi'(t))^{2}dt  \right\}  d\nu_{H}=\frac{\pi\sigma^{2}}{\pi^{2}-\alpha^{2}}\,\exp\left\{-\frac{2\left(\pi^{2} - \alpha^{2} \right)}{\sigma^{2}} \right\}\,.
$$

For $SL(2, \textbf{R})$-non-invariant integrands $ \Psi(h)$, the form of the singularity at $ \alpha\rightarrow\pi - 0$ may be different \cite{(BShCorrel)}, but in any case, the singularities in the nominator and in the denominator cancel each other in
 (\ref{JR}).

Now we briefly demonstrate how quasi-invariance of the measure is used for explicit evaluation of functional integrals (see  \cite{(BShCalculus)} for details).
 As a first step, it is convenient to represent functional integrals over the group $Diff^{1}_{+} (S^{1})$ as those over the group $Diff^{1}_{+}([0,1])$ with the ends of the interval $[0,\,1]$ glued $(\varphi'(0)=\varphi'(1))\,.$
In \cite{(BShCorrel)}, we proved
 the following equation:
\begin{equation}
   \label{Equality}
\frac{1}{\sqrt{2\pi}\sigma}\int\limits_{Diff^{1}_{+} (S^{1}) }F(h)\mu_{\sigma}(dh)
=\int\limits_{Diff^{1}_{+} ([0,1]) }\delta\left(\frac{h'(1)}{h'(0)}-1 \right)\,F(h)\,\mu_{\sigma}(dh)
\,.
\end{equation}

The quasi-invariance of the measure  (\ref{MeasureS}) with respect to the left action of the subgroup  $Diff^{3}_{+}([0,\,1])$
$\,(g\circ h\,,\ \ g\in Diff^{3}_{+}([0,\,1])\,,\ \ h\in Diff^{1}_{+}([0,\,1])\,)\,$ is written  \cite{(Shavgulidze1978)}, \cite{(Shavgulidze2000)} as
  $$
  \int \limits _{Diff^{1}_{+}([0, 1])}\,F(\tilde{h})\mu_{\sigma}(d\tilde{h})=\frac{1}{\sqrt{g'(0)g'(1)}}\int \limits _{Diff^{1}_{+}([0, 1])}\,F\left(g(h)\right)
  $$
  \begin{equation}
   \label{FI1}
  \times \exp\left\{ \frac{1}{\sigma^{2}}\left[   \frac{g''(0)}{g'(0)}h'(0)-  \frac{g''(1)}{g'(1)}h'(1)\right]   +        \frac{1}{\sigma^{2}}\int \limits _{0}^{1}\, \mathcal{S}ch\,\{g,\,h(t)\}\,\left(h' (t)\right)^{2}\,dt  \right\} \,\mu_{\sigma}(dh)\,.
\end{equation}

Let the function $g$  be
\begin{equation}
   \label{g}
g(t)=g_{\alpha}(t)=\frac{1}{2}\left[ \frac{1}{\tan\frac{\alpha}{2}}\tan\left(\alpha(t-\frac{1}{2}) \right)+1 \right]\,.
\end{equation}

In this case,
\begin{equation}
   \label{arc}
\left(g_{\alpha}^{-1}(h)\right)(t)=\frac{1}{\alpha}\arctan\left[\tan \frac{\alpha}{2}\,\left(2h(t)-1\right)\right]+\frac{1}{2}\,.
\end{equation}

And as the result of (\ref{FI1}), $J^{\alpha}$
is transformed into the integral
\begin{equation}
   \label{FI4}
\frac{\alpha}{\sin \alpha}
\,\int \limits _{Diff^{1}([0, 1])}\exp\left\{\frac{4\,\sin ^{2}\frac{\alpha}{2}}{\sigma^{2}}\left(h'(0)+h'(1)\right)\right\}
\,\Psi \left(g^{-1}_{\alpha}(h)\right)\,\delta\left(\frac{h'(1)}{h'(0)}-1 \right)\,\mu_{\sigma}(dh)
\end{equation}
that can be reduced to ordinary integrals by the method described in \cite{(BShCalculus)}.

\section {Factorization of the measure on $Diff^{1}_{+}(S^{1}) $}
\label{sec:factor}

In this section, we factorize the measure on $Diff^{1}_{+}(S^{1}) $
\begin{equation}
   \label{FactMeasure}
   \tilde{\mu}_{\sigma}(dh)=\nu _{H}(d\varphi)\ \tilde{\mu}_{\sigma}^{X}(df) \,,
\end{equation}
and hence demonstrate that, as the space to integrate over,
$Diff^{1}_{+}(S^{1}) $ is equivalent to the Cartesian product
$$
Diff^{1}_{+}(S^{1})\cong\,SL(2,\textbf{R})\,\times\,X\,,\ \ \ \ \ \ \
X=Diff^{1}_{+}(S^{1})/SL(2,\textbf{R})\,.
$$

First,  we take 3 points on the circle $t_{k}=\frac{k}{3}\,,\ \ k=0,\,1,\,2\,, $ and represent the integral (\ref{FI3}) as
\begin{equation}
   \label{FI3tau}
\frac{1}{\sqrt{2\pi}\sigma}\,\int\,d\tau_{0}\,d\tau_{1}\,d\tau_{2}\int\limits_{Diff^{1}_{+} (S^{1}) }\Psi(h)\prod\limits_{k=0,\,1,\,2}\delta\left(h(t_{k})-\tau_{k} \right)\,
 \tilde{ \mu}_{\sigma} (dh)\,.
\end{equation}

Now
we fix the 3 parameters of $SL(2,\textbf{R})$ assuming $\varphi_{\tau}(t_{k})=\tau_{k}\,,$ with $\tau_{k}$ being fixed.
Using the equality
$$
\delta\left(\varphi_{\tau}(f(t_{k}))-\tau_{k} \right)=\frac{1}{\varphi'_{\tau}(t_{k})}\delta\left(f(t_{k})-t_{k} \right)\,,
$$
we transform the integral (\ref{FI3tau}) into the form
\begin{equation}
   \label{FI3f}
\frac{1}{\sqrt{2\pi}\sigma}\int\prod\limits_{i=0,\,1,\,2}\frac{d\tau_{i}}{\varphi'_{\tau}(t_{i})}\int\limits_{Diff^{1}_{+} (S^{1});\, SL(2,R) gauge fixed }\Psi(\varphi_{\tau}\circ\ f)\prod\limits_{k=0,\,1,\,2}\delta\left(f(t_{k})-t_{k} \right)\,
 \tilde{ \mu}_{\sigma} (df)\,.
\end{equation}

It is convenient to consider the following realization of $SL(2,\textbf{R})$
\begin{equation}
   \label{SLrealiz}
\varphi_{\tau}(t)=\frac{1}{\pi}arc\cot\left\{A\cot(\pi t-\pi\theta)+B \right\}
\end{equation}
with the 3 parameters $A,\ B,\ \theta  $ related to $\tau_{0},\,\tau_{1},\,\tau_{2}\,.$

In terms of the  parameters $A,\ B, \ \theta  $, the measure $\prod\limits_{i=0,\,1,\,2}\frac{d\tau_{i}}{\varphi'_{\tau}(t_{i})} $ in (\ref{FI3f}) is written as
$$
const\, \frac{dB\,dA}{A^{2}}\,d\theta
$$
that is the Haar measure of $SL(2,\textbf{R})\ \ \nu_{H}(d\varphi)$ \cite{(Lang)}.

Thus, the integral (\ref{FI3f}) looks like
\begin{equation}
   \label{FI3f1}
\frac{1}{\sqrt{2\pi}\sigma}\int\limits_{ SL(2,R) }\nu_{H}(d\varphi)\,\int\limits_{Diff^{1}_{+} (S^{1});\, SL(2,R) gauge fixed }\Psi(\varphi_{\tau}\circ\ f)\prod\limits_{k=0,\,1,\,2}\delta\left(f(t_{k})-t_{k} \right)\,
 \tilde{ \mu}_{\sigma} (df)\,.
\end{equation}
And for $SL(2,\textbf{R})$ invariant integrands $\Psi(\varphi_{\tau}\circ\, f)=\Psi( f)\,, $ it is factorized
\begin{equation}
   \label{FI3f2}
\frac{1}{\sqrt{2\pi}\sigma}\int\limits_{ SL(2,R) }\nu_{H}(d\varphi)\,\int\limits_{Diff^{1}_{+} (S^{1})/ SL(2,R)}\Psi( f)\prod\limits_{k=0,\,1,\,2}\delta\left(f(t_{k})-t_{k} \right)\,
 \tilde{ \mu}_{\sigma} (df)\,.
\end{equation}

Note that we could fix the parameter $\tau_{0}$ (and correspondingly the parameter $\theta$) to be equal to zero at once from the very beginning. This choice
of the parameter fixes the point $0$ on the unit circle.

Now we turn to the integral over the quotient space
$$
\frac{1}{\sqrt{2\pi}\sigma}\int\limits_{Diff^{1}_{+} (S^{1})/ SL(2,R)}\Psi( f)\prod\limits_{k=0,\,1,\,2}\delta\left(f(t_{k})-t_{k} \right)\,
 \tilde{ \mu}_{\sigma} (df)
$$
\begin{equation}
   \label{FI3f3}
=\int\limits_{Diff^{1}_{+} ([0,\,1])}\Psi( f)\ \delta\left(f(\frac{1}{3})-\frac{1}{3} \right)\,\delta\left(f(\frac{2}{3})-\frac{2}{3} \right)\,\delta\left(\frac{f'(1)}{f'(0)}-1 \right)
 \tilde{ \mu}_{\sigma} (df)\,.
\end{equation}

We  split the interval $[0,\,1]$ into 3 parts $[0,\,\frac{1}{3}],\ \ [\frac{1}{3},\,\frac{2}{3}],\ \ [\frac{2}{3},\,1]$ and represent the function $f$ as
$$
f(t)=\frac{1}{3}f_{1}(3t)\,,\ \ 0\leq t \leq \frac{1}{3}\,;\ \ \ \
f(t)=\frac{1}{3}\left[1+f_{2}(3t-1)\right]\,,\ \ \frac{1}{3}\leq t \leq \frac{2}{3}\,;
$$
$$
f(t)=\frac{1}{3}\left[2+f_{3}(3t-2)\right]\,,\ \ \frac{2}{3}\leq t \leq 1\,.\ \ \ \ \ \
f_{1},\,f_{2},\,f_{3}\in  Diff^{1}_{+} ([0,\,1])\,.
$$

Using the technique
developed in \cite{(BShCalculus)} one immediately obtains
$$
\mu_{\sigma} (df)=\mu_{\frac{\sigma}{\sqrt{3}}} (df_{1})\ \mu_{\frac{\sigma}{\sqrt{3}}} (df_{2})\ \mu_{\frac{\sigma}{\sqrt{3}}} (df_{3})\,,
$$
and
$
\ \ \delta\left(f(\frac{1}{3})-\frac{1}{3} \right)=\frac{9}{2}f'_{1}(1)\ \delta\left(f'_{2}(0)-f'_{1}(1) \right)\,,\ \ \
\delta\left(f(\frac{2}{3})-\frac{2}{3} \right)=6f'_{2}(1)\ \delta\left(f'_{3}(0)-f'_{2}(1) \right)\,,
$
$$
\delta\left(\frac{f'(1)}{f'(0)}-1 \right)=f'_{3}(1)\ \delta\left(f'_{1}(0)-f'_{3}(1) \right)\,.
$$
Therefore, the integral (\ref{FI3f3}) is written as
$$
\int\limits_{Diff^{1}_{+} ([0,\,1])}\int\limits_{Diff^{1}_{+} ([0,\,1])}\int\limits_{Diff^{1}_{+} ([0,\,1])}\Psi( f_{1},\,f_{2},\,f_{3})\ 3^{3}\,
f'_{1}(1)\,f'_{2}(1)\,f'_{3}(1)
$$
$$
\times\,\delta\left(f'_{2}(0)-f'_{1}(1) \right)\,\delta\left(f'_{3}(0)-f'_{2}(1) \right)\,\delta\left(f'_{1}(0)-f'_{3}(1) \right)
$$
\begin{equation}
   \label{FI3f4}
\times\,\exp\left\{\frac{2\pi^{2}}{3\sigma^{2}}\int \limits _{0}^{1}\,\left[\left(f'_{1}(t)\right)^{2}+\left(f'_{2}(t)\right)^{2}+\left(f'_{3}(t)\right)^{2}\right]\,dt  \right\}\,
\mu_{\frac{\sigma}{\sqrt{3}}} (df_{1})\ \mu_{\frac{\sigma}{\sqrt{3}}} (df_{2})\ \mu_{\frac{\sigma}{\sqrt{3}}} (df_{3})\,.
\end{equation}

Functional integration in (\ref{FI3f4}) by the method based on quasi-invariance  \cite{(BShCalculus)}
obviously leads to the singularity-free result. (In this case, $\alpha =\frac{\pi}{3}\,, $ cf. (\ref{FI1})-(\ref{FI4})).

\section{Concluding remarks}
\label{sec:concl}

The explicit form of the measure on
 the quotient space $Diff^{1}_{+}(S^{1})/SL(2,\textbf{R})$
 $$
 3^{3}\
f'_{1}(1)\,f'_{2}(1)\,f'_{3}(1)
\ \delta\left(f'_{2}(0)-f'_{1}(1) \right)\ \delta\left(f'_{3}(0)-f'_{2}(1) \right)\ \delta\left(f'_{1}(0)-f'_{3}(1) \right)
$$
\begin{equation}
   \label{FI3f5}
\times\,\exp\left\{\frac{2\pi^{2}}{3\sigma^{2}}\int \limits _{0}^{1}\,\left[\left(f'_{1}(t)\right)^{2}+\left(f'_{2}(t)\right)^{2}+\left(f'_{3}(t)\right)^{2}\right]\,dt  \right\}\,
\mu_{\frac{\sigma}{\sqrt{3}}} (df_{1})\ \mu_{\frac{\sigma}{\sqrt{3}}} (df_{2})\ \mu_{\frac{\sigma}{\sqrt{3}}} (df_{3})\,.
\end{equation}
is obtained in the present paper for the first time. (See, e.g., \cite{(SW)}, p. 9, footnote 7.)

The paper complements the previous studies of the rules of  Schwarzian functional integration  \cite{(BShExact)} -  \cite{(BShCalculus)}.
With the regular technique of evaluation of functional integrals developed there, the result obtained in the paper gives new possibilities for studying a wide class of theories.

\end{document}